\documentclass[%
 aip,
 jmp,%
 amsmath,amssymb,
reprint,%
]{revtex4-1}

\usepackage{graphicx}
\usepackage{dcolumn}
\usepackage{bm}
\usepackage{hyperref}
\usepackage[usenames,dvipsnames,svgnames,table]{xcolor}

\begin{document}

\preprint{AIP/123-QED}

\title{Ultrahigh Q-Frequency product for optomechanical disk resonators with a mechanical shield}

\author{D. T. Nguyen}
\email{dac-trung.nguyen@univ-paris-diderot.fr}
\author{C. Baker}
\author{W. Hease}
\author{S. Sejil}
\affiliation{Laboratoire Mat\'eriaux et Ph\'enom\`enes Quantiques, Universit\'e Paris Diderot -- CNRS, 10 rue Alice Domon et L\'eonie Duquet, 75013 Paris, France}

\author{P. Senellart}
\author{A. Lema\^itre}
\affiliation{Laboratoire de Photonique et de Nanostructures, route de Nozay, 91460 Marcoussis, France}

\author{S. Ducci}
\author{G. Leo}
\author{I. Favero}
\affiliation{Laboratoire Mat\'eriaux et Ph\'enom\`enes Quantiques, Universit\'e Paris Diderot -- CNRS, 10 rue Alice Domon et L\'eonie Duquet, 75013 Paris, France}

\date{\today}

\begin{abstract}

We report on optomechanical GaAs disk resonators with ultrahigh quality factor - frequency product $Q\cdot f$. Disks standing on a simple pedestal exhibit GHz breathing modes attaining a $Q\cdot f$ of $10^{13}$ measured under vacuum at cryogenic temperature. Clamping losses are found to be the dominant source of dissipation in this configuration. A new type of disk resonator integrating a shield within the pedestal is then proposed and its working principles and performances investigated by numerical simulations. For dimensions compatible with fabrication constraints, the clamping-loss-limited $Q$ reaches $10^{7}-10^{9}$ corresponding to $Q\cdot f$ of $10^{16}-10^{18}$. This shielded pedestal approach applies to any heterostructure presenting an acoustic mismatch.

\end{abstract}


\maketitle

The interaction of light with mechanical motion - optomechanics \cite{marquardt_optomechanics_2009, favero_optomechanics_2009, aspelmeyer_cavity_2013} and its related concepts - is now investigated in a wide variety of experimental settings. Optomechanical resonators of various size and geometry continue to be developed and optimized for applications like weak force sensing \cite{forstner_cavity_2012, miao_microelectromechanically_2012} or optical cooling of mesoscopic mechanical systems down to the quantum regime \cite{teufel_sideband_2011, chan_laser_2011}. For most of these applications, high mechanical frequency f, strong optomechanical coupling $g_0$ and low optical/mechanical dissipation are desirable.
Among various systems, Gallium Arsenide (GaAs) optomechanical disk resonators bring together all of these features with a relatively simple geometry \cite{ding_wavelength-sized_2011} and the possibility of complete on-chip optical integration \cite{baker_critical_2011}. The sub-micron optical and mechanical confinement leads to GHz mechanical frequencies and $g_0$ in the MHz \cite{ding_high_2010}.  Optical quality factors reach today several $10^{5}$ and optical dissipation sources are progressively unraveled to approach the radiative limit. On the mechanical dissipation side, the understanding of loss mechanisms in GaAs disks is largely incomplete, despite recent efforts to model their fluid damping for air or liquid operation \cite{parrain_damping_2012}.

Mechanical resonators are often compared on the basis of their mechanical $Q\cdot f$ factor (quality factor times frequency) because of the paramount importance of this figure of merit for the performances of MEMS devices \cite{nguyen_mems_2007}. But $Q\cdot f$ also turns out to play a key role in the quantum realm, where it indicates the number of independent operations $N$ that can be performed on a quantum mechanical system subject to thermal decoherence induced by an environment at temperature $T$ \cite{devoret_nanomechanical_2012}. More specifically, the number of coherent oscillations in presence of a thermal environment is given by $Q\cdot f \times h/(k_\text{B} T)$, which indicates that a $Q\cdot f$ higher than $6\cdot 10^{12}$ is necessary to attain one coherent oscillation at room temperature. Two independent works have demonstrated record values for $Q\cdot f$ in the $10^{15}-10^{16}$ range for quartz resonators at ultra-low temperature \cite{goryachev_extremely_2012, smagin_1974} and very recent developments on Silicon optomechanical crystals  allowed reaching a $Q\cdot f$ of $10^{14}$ \cite{chan_laser_2011}. Apart from these three works, current state of the art systems are evolving in the $10^{10}-10^{13}$ window (see Ref. \cite{devoret_nanomechanical_2012} and \cite{aspelmeyer_cavity_2013} for more comprehensive reviews on this topic) and are based on Quartz, Silicon, and Polysilicon, Silicon Nitride or Diamond, and with very few reports on III-V semiconductors. First studies on GaAs disks reported $Q\cdot f$ products between $10^{11}$ and $10^{12}$ in ambient conditions \cite{ding_high_2010, ding_wavelength-sized_2011}, at the forefront of GaAs based mechanical systems \cite{liu_high-q_2011, mahboob_bit_2008, cole_monocrystalline_2008}. 

In this letter we focus on mechanical dissipation and the $Q\cdot f$ factor in GaAs disks, with an emphasis on clamping losses. By measuring and modeling the mechanical $Q$ of disks of varying pedestal radius, we find that clamping loss is the dominant loss mechanism when these resonators sit on a simple central pedestal and are operated in vacuum at low temperature. As compared to our previous work, the improved control of the pedestal fabrication allows miniature GaAs disk resonators reaching a $Q\cdot f$ factor of $10^{13}$. Building on the obtained numerical understanding of clamping losses, we propose a mechanical shield geometry that allows boosting the clamping limited $Q$ from some thousands in the simple-pedestal geometry to the $10^{7}-10^{9}$ range with the shield, corresponding to a clamping-limited $Q\cdot f$ factor of $10^{16}-10^{18}$. We present an optimization of this shield geometry, notably under fabrication constraints, and  give a physical discussion of the decoupling of the disk from its support in this geometry.

The disk samples are fabricated from a GaAs (320 nm)/Al$_{0.8}$Ga$_{0.2}$As (1800 nm) bilayer grown by molecular beam epitaxy (MBE) on a GaAs substrate. Relevant mechanical properties of GaAs and Al$_{0.8}$Ga$_{0.2}$As seen as isotropic elastic materials are summarized in Table I. 

\begin{table}[htbp]
\centering
{\renewcommand{\arraystretch}{1.5}
\renewcommand{\tabcolsep}{0.1cm}
\begin{tabular}[c]{lccc}
\hline
%\vspace{-0.2cm}\\
\textbf{Parameter} & \textbf{Unit}  & \textbf{GaAs} & \textbf{Al$_{0.8}$Ga$_{0.2}$As}\\
\hline
%\vspace{-0.2cm}\\
Young's modulus ($E$) & GPa & 85.9 & 83.9\\
Density ($\rho$) & kg/m$^{3}$ & 5317 & 4072\\
Poisson's ratio ($\nu$) & - &  0.310 & 0.318\\
\hline
\end{tabular}}
\caption{Mechanical properties of GaAs \cite{comsol_library} and Al$_{0.8}$Ga$_{0.2}$As \cite{adachi_gaas_1985, adachi_properties_1993}.\label{tab:tab1}}
\end{table}

\begin{figure}[htbp]
\centering
\includegraphics[width=\columnwidth]{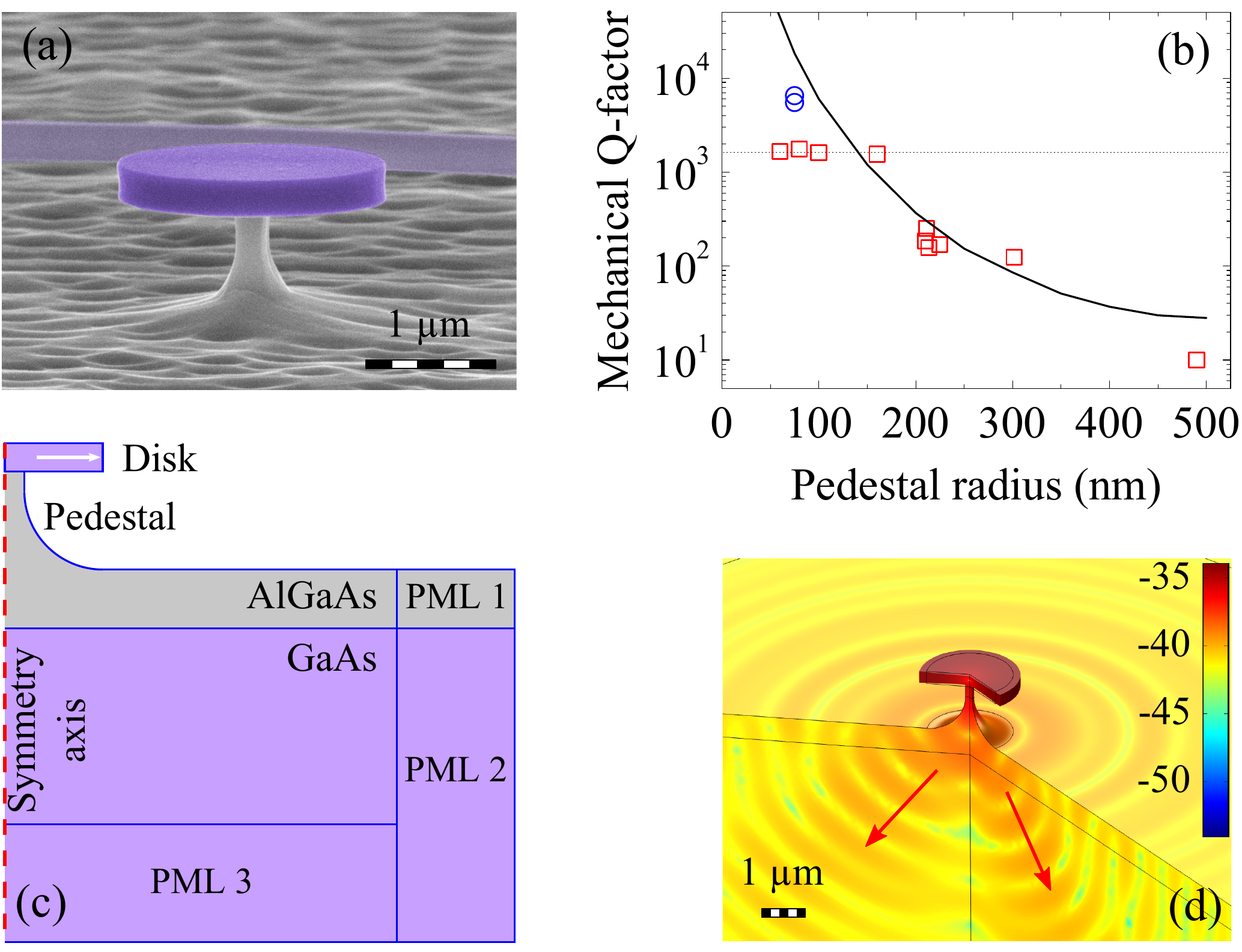}
\caption{GaAs disks with a simple pedestal. (a): SEM side-view of a GaAs disk and its coupling optical waveguide in the background (in purple false color). The waveguide is 200 nm wide and has the same thickness of the disk. Roughness on the ground is due to HF wet etching of AlGaAs (in gray). (b) Experimental data (open red squares corresponding to values measured in air at 300 K and blue circles corresponding to values measured in vacuum at 8 K) and numerical simulations (black solid line) for the mechanical $Q$ of disk radial breathing mode as a function of pedestal radius. (c) Mechanical modeling of a GaAs disk resonator. The driving force used to simulate the disk's spectral response is a uniform pressure acting on the disk's sidewall (white arrow). (d) Simulated instantaneous displacement field (in log color scale). Red arrows show the propagation of the deformation wave.\label{fig:fig1}}
\end{figure}

Disks of radius 1 $\mu$m are positioned in the vicinity of tapered suspended GaAs waveguides to allow evanescent optical coupling of light into the disks. They are patterned in a resist mask by electron beam lithography and then dry-etched by non-selective Inductively Coupled Plasma Reactive Ion Etching (ICP-RIE) using a mixture of SiCl$_4$ and Ar plasmas. Pedestals are formed by hydrofluoric acid (HF) selective underetching of the AlGaAs sacrificial layer. A diluted HF:H$_2$O solution (1.22 \% in volume) at 4$^\circ$C is combined with a slow agitation in the solution to allow fabricating disks with a controlled pedestal radius as small as 60 nm. Protecting the AlGaAs parts from air oxidation by putting the sample in acetone right after ICP-RIE proved crucial to obtain the degree of control needed in the present work. Fig. \ref{fig:fig1}a shows a finished GaAs disk and its coupling waveguide with smooth sidewalls.

We first experimentally measure the mechanical spectrum of several GaAs disk resonators. Optical probing with a laser blue-detuned to an optical whispering-gallery resonance gives access to the mechanical spectrum. By virtue of the optomechanical coupling, the disk mechanical fluctuations are imprinted in the optical transmission noise and analyzed at the photodetector output using a spectrum analyzer. Details of the experiment setup can be found in our previous work \cite{ding_high_2010, baker_critical_2011, ding_wavelength-sized_2011}. In the low laser power limit, the obtained spectra provide the intrinsic frequencies and quality factors of the disk mechanical modes. We focus here on the mechanical breathing mode of the disks, which appears around 1.37 GHz for the considered dimensions. Experiments are run both at room temperature and at 8 K. Measurements reveal that a reduction of pedestal radius from 500 to 100 nm barely lowers the mechanical frequency (less than 5\%, not shown here). The measured $Q$ factors are presented in Fig. \ref{fig:fig1}b and show a considerable 15-fold increase between 300 and 70 nm of pedestal radius at room temperature in air (red squares). Under vacuum operation at low temperature (8 K), a $Q$ of $6500 \pm 1000$ is measured for the smallest investigated pedestal radius (blue circles).  This corresponds to the highest $Q\cdot f$ value reported in GaAs disk resonators, attaining the $10^{13}$ limit. In air, $Q$ is limited at about 1700, probably due to air damping \cite{parrain_damping_2012}.

Experimental data are compared to finite-element method (FEM) simulations using the Comsol software. Fig. \ref{fig:fig1}c depicts a GaAs disk in a 2D axisymmetric model. To reflect the morphology resulting from fabrication, the disk is modeled by a cylinder and the pedestal by a cylinder followed by an isotropic etch profile. The disk and its support are standing on a substrate surrounded by perfectly matched layers (PML) that emulate the attenuation of the deformation wave. The dimensions of the substrate and PML parts are optimized to correctly simulate wave propagation within a manageable computing time. Fig. \ref{fig:fig1}d shows the first radial breathing mode (RBM) of a disk with a frequency around 1.37 GHz. This is the only mechanical mode detected in this frequency range in the experiments. Fig. 1d shows clearly the propagation of the deformation wave from the disk through the pedestal before dissipating in the substrate. The simulated mechanical $Q$ versus the pedestal radius is shown as a solid line in Fig. \ref{fig:fig1}b, which captures qualitatively and quantitatively the increase of $Q$ for smaller pedestal radius. The residual differences between experimental data and simulations can be ascribed to geometry imperfections and other dissipation channels such as surface-state loss which depend on the bath temperature. For example, we have observed a reduction of $Q$ to about 3000 (data not shown here) for the smallest radius disks as the temperature is increased from 8 to 300 K. Our experimental and numerical results indicate that support loss is the main dissipation channel of these GaAs disk optomechanical resonators at 8 K. As a consequence, a simple way to boost $Q$ is to reduce the impact of the anchoring points. A first natural route is to further reduce the pedestal radius, but this comes with two major drawbacks: Firstly, the system would become extremely fragile and secondly, thermal effects like thermo-optical instabilities would be exacerbated, given that the pedestal is also the main thermal dissipation channel.

Here we explore a second route inspired by phononic Bragg mirrors in order to prevent the disk deformation wave from escaping to the substrate and confine it in the resonator \cite{trigo_confinement_2002}. To form a periodic multilayered acoustic Bragg mirror, one needs two or more materials with different acoustic impedances, which are naturally GaAs and Al$_{0.8}$Ga$_{0.2}$As in our case. The phononic mirror could in principle be integrated within the substrate, just under the disk pedestal, or within the pedestal itself. However, Fig. \ref{fig:fig1}d reveals that the deformation wave becomes quasi-spherical as it exits the pedestal and propagates through the substrate. Hence a conventional planar Bragg mirror under the pedestal would not block the wave efficiently, as confirmed by our simulations (not shown). Therefore the Bragg structure must be integrated directly into the pedestal in order to prevent the wave from reaching the substrate. A standard Bragg mirror consists of quarter-wavelength ($\lambda$/4) layers. For a phonon mode at frequency of 1.37 GHz considered here, the  ``acoustic'' wavelength is $\lambda = 3.4$ $\mu$m in GaAs and 3.9 $\mu$m in Al$_{0.8}$Ga$_{0.2}$As. This implies that each layer has a thickness of about 1 $\mu$m. Besides, the first AlGaAs layer under the GaAs disk should be thick enough to minimize light coupling from the disk to the substrate in the final device, and thin enough to avoid growth and etching difficulties. This leads us to chose an optimal thickness of 2 $\mu$m. Finally, because of fabrication limitations, we avoid vertical etch depths of more than 10 $\mu$m and therefore focus on structures with a small number of layers.

\begin{figure}[ht]
\centering
\includegraphics[width=\columnwidth]{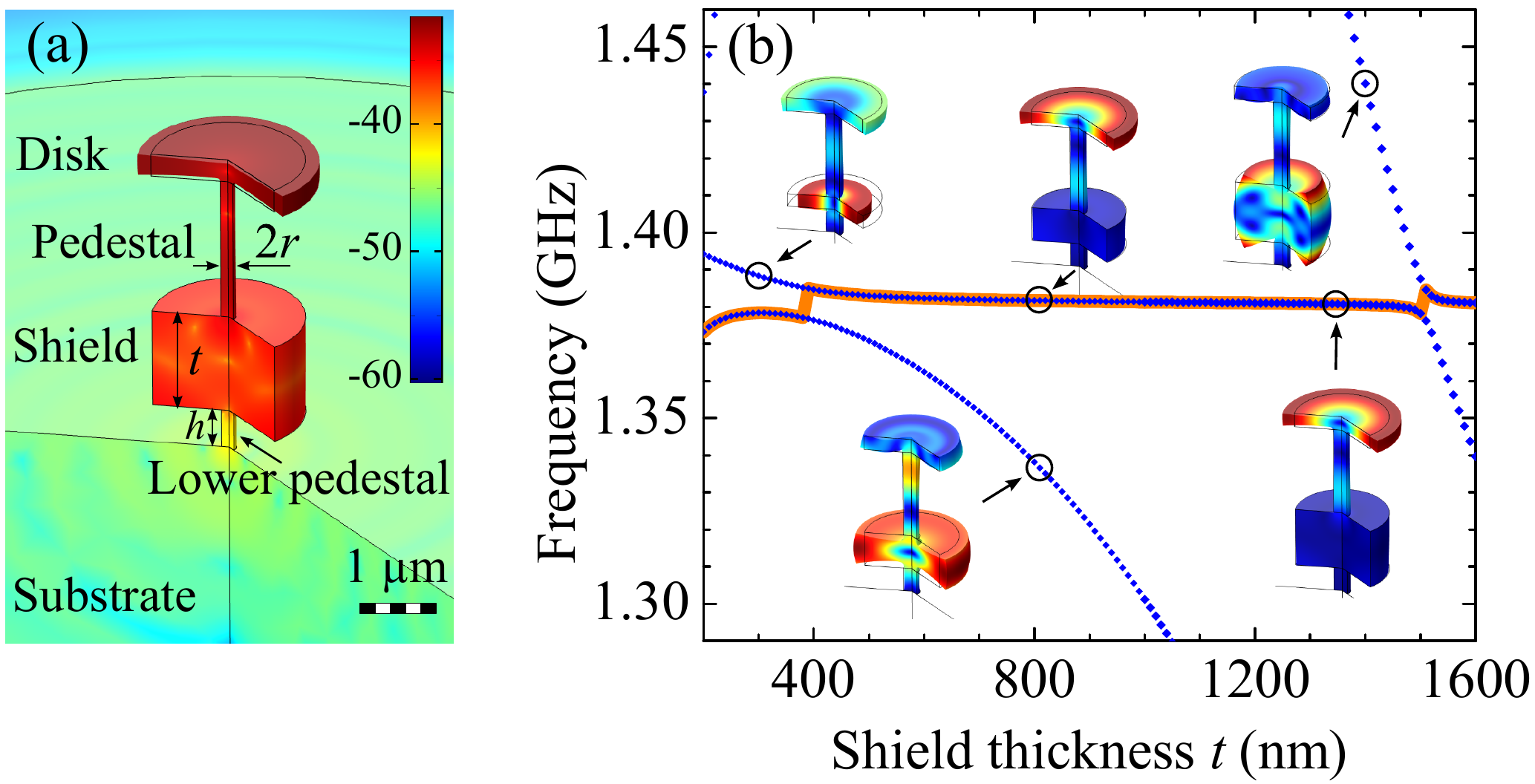}
\caption{Shielded disk resonator. (a) Geometry of the shielded disk. Disk and shield are both 1 $\mu$m in radius. Color log scale represents the displacement field for a breathing mode at 1.37 GHz. Notice the high color contrast between the disk and substrate showing excellent isolation of mechanical vibration. (b) Mechanical mode dispersion as a function of the shield thickness. Blue dotted lines represent mechanical modes. The bold red line points towards modes with the highest disk radial displacement amplitude. Pictures show the displacement profile of selected modes.\label{fig:fig2}}
\end{figure}

These constraints lead us to the structure shown in Fig. 2a. It consists of a 320 nm thick GaAs disk on a first AlGaAs pedestal of 2 $\mu$m in length, that stands on a GaAs ``shield'', the latter topping a second (lower) AlGaAs pedestal to isolate the shield from the GaAs substrate. This structure can be fabricated using the same techniques as for the disks with simple pedestal. As a result, the disk and the shield will have the same radius of 1 $\mu$m and the upper and lower pedestals share a common radius. The structure's mechanical properties are computed numerically as in the simple-pedestal case above (i.e. the disk, the pedestals and the shield are modeled by cylinders in a 2D axisymmetric approach). The adjustable dimensions for optimization are the pedestal radius $r$, the shield thickness $t$ and the height $h$ of the lower pedestal. Compared to a simple disk presented in Fig. \ref{fig:fig1}c, the higher color contrast between the disk and the substrate in Fig. \ref{fig:fig2}a  illustrates the good efficiency of the shield at confining the mechanical RBM mode within the disk.

\begin{figure}[ht]
\centering
\includegraphics[width=\columnwidth]{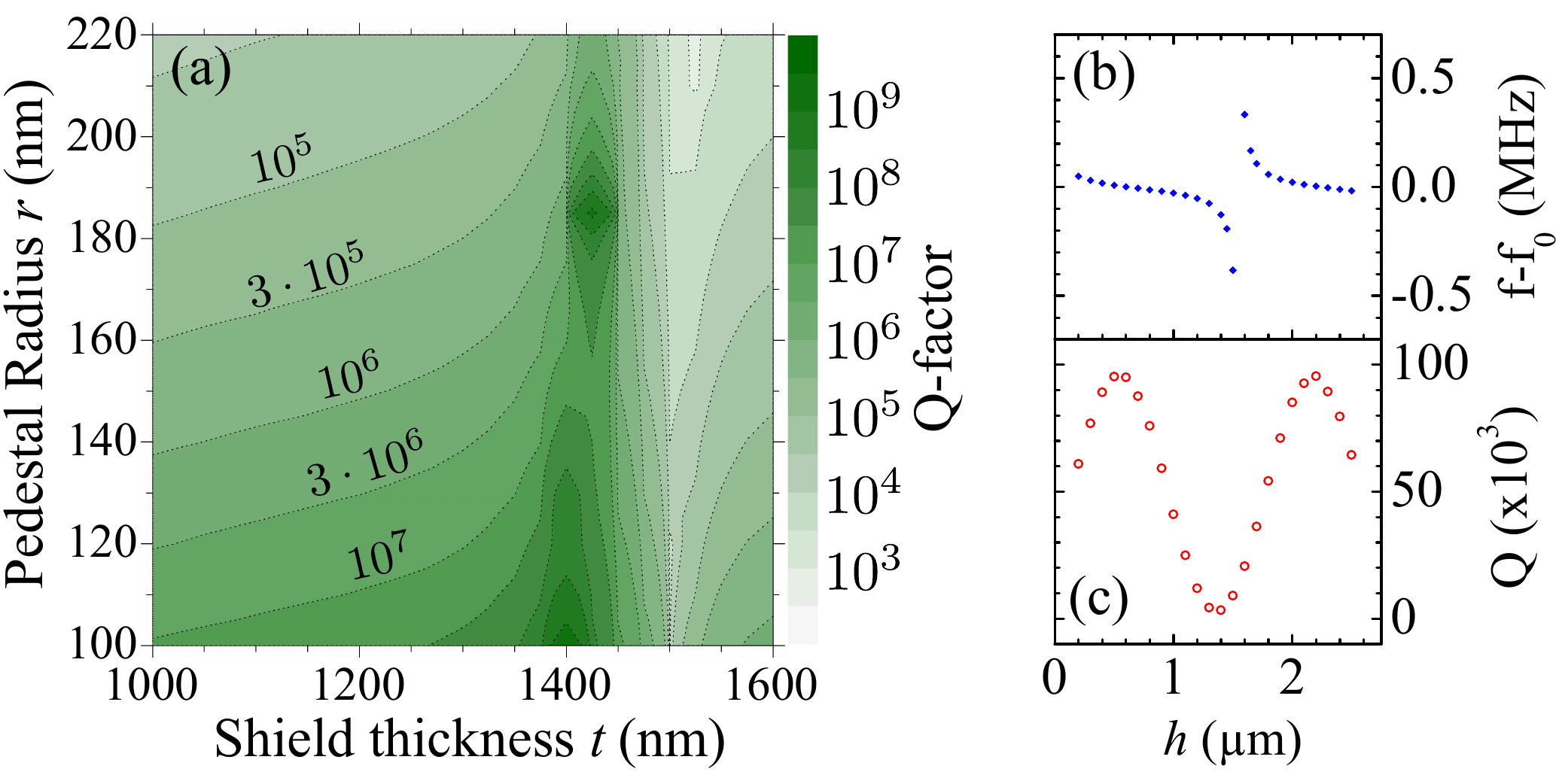}
\caption{$Q$ optimization of the geometry presented in Fig. \ref{fig:fig2}a. (a) Contour color map in log scale of the calculated $Q$-factor of a shielded disk resonator as a function of shield thickness and pedestal radius for $h = 550$ nm. (b) Radial breathing mode frequency and $Q$-factor as a function of the lower pedestal's height $h$, with $(t;r)=(1000;180)$ nm. Frequency scale is offset by $f_0=1.38$ GHz for clarity.\label{fig:fig3}}
\end{figure}

Fig. \ref{fig:fig3}b shows the computed mechanical modes of the shielded resonator in the frequency range of the radial breathing mode considered above (1.30-1.45 GHz). As the shield thickness varies from 200 to 1600 nm, three distinct mechanical modes appear, each of them represented by a blue dotted dispersion line. The pedestal radius and the lower pedestal height are respectively fixed to 180 nm and 550 nm in these simulations ($h=550$ nm is optimal as discussed below). As apparent in the displacement profiles shown in the figure, normal modes of the structure result mainly from a coupling between eigenmodes of the disk and the shield. Each mode shows a disk radial breathing nature in a given shield thickness range highlighted by  the red line. The radial breathing mode of the disk is hybridized with eigenmodes of the shield and the details of this hybridization will be important to minimize support losses.

To determine the optimal geometry with minimal support losses and highest $Q$ we proceed as follows: 1) Study the dependence of $Q$ on the shield thickness $t$ and pedestal radius $r$ by fixing $h$ at a certain value. 2) Analyze the dependence of $Q$ on $h$ for several values of $(t;r)$. This step reveals that the optimal $h=550$ nm is independent of the selected $(t;r)$ in our investigation. 3) Finally repeat the first step with the optimal value of $h$. This strategy leads to the results shown in Fig. \ref{fig:fig3}.

The clamping-limited mechanical $Q$ of the disk GHz breathing mode can reach extremely large values. In Fig. \ref{fig:fig3}a, a first noticeable region of the parameter space around $(t;r) = (1425;185)$ nm provides for example $Q = 10^7-10^9$. This configuration is however not the best technological compromise because of the related narrow tolerances on $t$ and $r$ and the rapid drop of $Q$ if $t$ becomes too large. To account for some 1\% imprecision on the epitaxial thickness $t$ and for the realistic pedestal etching conditions - obtaining a 10 nm precision on $r$ in the region $r\sim 150$ nm remains challenging - we estimate that a good compromise is reached in another region of the parameter space (with $1350\leq t \leq 1400$ nm and $r\leq 200$ nm) that also provides ultrahigh $Q$ from $3\cdot 10^5$ to $10^9$, but with larger tolerances on the fabrication. 

Even though our shielded disk geometry is originally inspired by planar Bragg mirrors, its dimensions and working principles are somewhat different due to the small number of employed layers and their finite lateral size. The boost of $Q$ in shielded resonators can be understood in terms of interference between deformation waves emitted by the disk and the shield. In case of $Q$ enhancement, the interference is destructive in the lower pedestal, which results in a better isolation from the substrate. In this situation the disk and shield oscillate in anti-phase: when the disk expands, the shield contracts. The disk's expansion pulls the rest of the structure towards the disk while the shield contraction pushes away. These two actions add up constructively in the top parts of the structure (above the shield) but cancel out at the lower pedestal. For specific values of the shield thickness t this cancellation is quasi-total, resulting in ultrahigh $Q$. This interpretation as an interference between different modes is corroborated by the last results shown in Fig. \ref{fig:fig3}. Fig. \ref{fig:fig3}b indicates an anti-crossing as the longitudinal extensional mode of the lower pedestal approaches the disk breathing mode in frequency. In the vicinity of this anti-crossing a drop in $Q$ is observed \cite{anetsberger_ultralow-dissipation_2008, sun_high-q_2012}. A more complete sinusoidal dependence of $Q$ on $h$ is shown in Fig. \ref{fig:fig3}c and indicates the crucial role of the standing wave formation in the lower pedestal. A high (low) $Q$ value corresponds to the case where the anchoring point to the substrate is located at a node (an antinode) of this wave.

In summary, we have reported what is to our knowledge the highest $Q\cdot f$ value for a GaAs mechanical resonator. Numerical simulations show that clamping losses are the dominant source of mechanical dissipation in this generation of resonators standing on a simple pedestal. In order to quench this source of loss, we propose a shielded disk geometry whose first radial breathing mode at $\sim 1.38$ GHz is expected to attain a clamping-limited $Q$ as high as $10^{9}$ corresponding to $Q\cdot f = 10^{18}$. A further advantage of the proposed shielded geometry is its simplicity, which makes it applicable to any resonator built from a hetero-structured wafer with some acoustic mismatch. III-V semiconductors naturally lend themselves to these ideas and could allow to connect mechanical modes of ultimately low dissipation with active photonic elements like a single quantum dot or quantum well, opening the exploration of hybrid cavity-QED and optomechanics scenarios in semiconductors \cite{restrepo_single_2013}.

This work is supported by the French ANR through the NOMADE project and by the ERC through the GANOMS project. The authors thank E. Gil-Santos for technical hints.

\bibliographystyle{ieeetr}

\end{document}